\documentstyle[prl,aps,multicol,epsf]{revtex}
\tighten
\begin{document}
\draft

\title{Electron spectroscopy of the shot noise reduction effect}
\author{O. M. Bulashenko and J. M. Rub\'{\i}}
\address{Departament de F\'{\i}sica Fonamental,
Universitat de Barcelona, Diagonal 647, E-08028 Barcelona, Spain}
\author{V. A. Kochelap}
\address{Department of Theoretical Physics,
Institute of Semiconductor Physics, Kiev 252028, Ukraine}
\date{June 5, 1999}
\maketitle

\begin{abstract}
A general formula for current noise in a two-terminal ballistic nondegenerate 
conductor under the action of long-range Coulomb correlations has been derived.
The noise reduction factor (in respect to the uncorrelated value)
is obtained for biases ranging from thermal to shot noise limits, and 
it is related to spatial variations in transport characteristics.
The contributions of different energy groups of carriers to the noise 
are found, that leads us to suggest an electron energy spectroscopy experiment 
to probe the Coulomb correlations in ballistic conductors.
\end{abstract}

%\pacs{PACS numbers: 72.70.+m, 73.23.Ad, 73.50.Td, 05.40.Ca}

\begin{multicols}{2}

Shot-noise measurements become a fundamental tool to probe carrier 
interactions in mesoscopic systems \cite{landauer98}.
The term ``shot noise'', appeared originally in the context of pure
ballistic electron transmission in vacuum-tube devices, has acquired
nowadays a much broader usage and refers to different mesoscopic structures,
including diffusive conductors, resonant-tunneling devices, etc., where
the carrier flow exhibits nonequilibrium noise proportional to
the electric current \cite{dejong97}. 

In this Letter we address the case of shot noise in pure ballistic conductors
under the action of long-range Coulomb correlations,
thus going back to the genuine shot noise definition.
The fact that shot noise may be affected by Coulomb interactions has been known
since the times of vacuum tubes \cite{vacuum}.
Their importance in mesoscopic conductors has been emphasized
by Landauer \cite{landauer} and B\"uttiker \cite{buttiker96},
and evidenced recently by Monte Carlo simulations \cite{gonzalez97,b98}.
The purpose of our paper is twofold.
First, we present for the first time a self-consistent theory of shot noise 
in ballistic nondegenerate conductors \cite{remark1} by solving analytically 
the kinetic equation.
Second, basing upon this theory, we suggest an electron spectroscopy 
experiment to make the Coulomb correlations effect visible.
The possibility of such an experiment is based on recent advances
in nanoscale fabrication techniques and shot-noise measurements
\cite{reznikov95,kumar96,schoel97}.

Consider a two-terminal semiconductor sample with plane parallel heavily-doped 
contacts at $x=0$ and $x=d$. 
We assume $\lambda_w\ll d\lesssim \lambda_p$, with $\lambda_w$ the electron 
wavelength and $\lambda_p$ the mean free path, 
so that electrons may be considered as 
classical particles moving ballistically between the contacts and 
interacting with each other electrostatically.
The transport in such a system is described by the Vlasov system of
equations, that is the collisionless kinetic equation
for the electron distribution function $F(X,v_x,t)$ coupled
self-consistently with the Poisson equation for the electrostatic potential
$\varphi(X,t)$

\begin{eqnarray} \label{vlas}
\frac{\partial F}{\partial t}
&+& v_x \frac{\partial F}{\partial X} +
\frac{q}{m}\frac{d\varphi}{dX} \frac{\partial F}{\partial v_x} = 0, \\
\frac{d^2\varphi}{dX^2} &=& \frac{q}{\kappa} \int F(X,v_x,t) dv_x.
\label{pois}
\end{eqnarray}
Here, $v_x$ is the $X$-component of the electron velocity,
$q$ the electron charge, $m$ the electron effective mass,
and $\kappa$ the dielectric permittivity.
The applied bias $U$ between the contacts is assumed to be fixed by
a low-impedance external circuit.
The distribution functions at the left (L) and right (R) contacts
are supposed to consist of a stationary part and a small fluctuation

\begin{eqnarray} \label{bcvlas}
F(0,v_x,t)|_{v_x>0} &=& F_s(v_x) + \delta F_L(v_x,t), \nonumber\\
F(d,v_x,t)|_{v_x<0} &=& F_s(v_x) + \delta F_R(v_x,t).
\end{eqnarray}
Under nondegenerate and equilibrium conditions in the contacts, 
we assume for the stationary part of the injection function
the half-Maxwellian distribution
$F_s(v_x)=[2 N_0/(v_0\sqrt{\pi})]e^{-v_x^2/v_0^2}$, where
$N_0$ is the density of electrons injected from the contact,
$v_0=\sqrt{2k_B T/m}$ is the thermal velocity, $k_B$ the Boltzmann constant,
and $T$ the temperature.
The stochastic terms $\delta F_k$, $k=L,R$ in Eq.\ (\ref{bcvlas}) are
the only sources of noise under ballistic transport considered here, since
the electron motion between the contacts is noiseless.
Their correlation is given by
$\langle\delta F_k(v_x,t)\delta F_{k'}(v_x',t')\rangle
= F_s(v_x)\delta_{kk'}\delta(v_x-v_x')\delta(t-t')$.
As a consequence of these fluctuations inside the contacts
(whose origin is ultimately the carrier scattering processes),
both the electron distribution function and electrostatic potential
in the ballistic sample fluctuate, which leads to the current fluctuations.
Introducing the Fourier transform for the fluctuations,
the linearized Vlasov equations give

\begin{eqnarray} \label{vlasf}
\left( v_x \frac{\partial}{\partial X}\right. &+& \left.
\frac{q}{m} \frac{d\varphi}{dX} \frac{\partial}{\partial v_x} \right) 
\delta F(X,v_x)
= - \frac{q}{m} \frac{\partial F}{\partial v_x} \frac{d\delta\varphi}{dX}, \\
\frac{d^2 \delta\varphi }{dX^2} &=&
\frac{q}{\kappa} \int \delta F(X,v_x) d v_x,\label{dpois}
\end{eqnarray}
where in the low-frequency regime of interest here ($\omega\ll\tau_T^{-1}$,
with $\tau_T$ being an average transit time between the contacts)
we have omitted the term $\propto i\omega$.
It is seen, that the calculation of fluctuations requires the knowledge of
the stationary distributions $F$ and $\varphi$, which, in turn,
can be determined by solving the self-consistent steady-state problem.
It is advantageous to introduce the dimensionless potential
$\psi(x)=q\varphi(x)/(k_B T)$, and to scale
$n = N/(2 N_0)$, $x=X/L_D$, with $L_D=\sqrt{\kappa k_B T/(2q^2 N_0)}$
being the Debye screening length.
In such units our problem contains only two dimensionless parameters:
(i) the length of the sample $\lambda=d/L_D$, and
(ii) the applied voltage bias $V=qU/(k_BT)$.
Let the space charge be such, that a potential minimum
$\psi_m\equiv -V_m$ occurs at $x=x_m$, which acts as a barrier
for the electrons by reflecting a part of them back to the contacts.
Since Eq.\ (\ref{vlas}) is equivalent to $(d F/d t)|_{trajectory} = 0$,
the distribution function at any plane $X$ may be expressed 
through the injection distribution functions given at the contacts. 
By making use of the Maxwellian stationary injection and the contribution 
of different groups of carriers (transmitted and reflected),
we obtain the electron density $n=\int F(x,v_x) dv_x$ as a functional
of the potential

\begin{equation} \label{den}
n(\eta)=n_m e^{\eta}\,
[1 \pm \beta \,{\rm erf}(\sqrt{\eta})],
\end{equation}
where $\eta(x)=\psi(x)-\psi_m$ is the shifted potential measured from
the minimum, erf stands for the error function,
$n_m={1\over 2}e^{-V_m}(1+e^{-V})$ is the electron density at the potential
minimum, and $\beta=\tanh(V/2)$.
Here, and throughout of the paper, we shall use the upper sign for the left
side of the potential minimum $0<x<x_m$, 
and the lower sign for the right side $x_m<x<\lambda$.
Note that in equilibrium, $V$=0, $\beta$=0, the Boltzmann distribution
$n(x) = e^{\psi(x)}$ is recovered throughout the sample.

The obtained Eq.\ (\ref{den}) is then used to solve the Poisson
equation $d^2\eta/dx^2=n$, subject to the boundary conditions
at the contacts $\eta(0)=V_m$, $\eta(\lambda)=V_m+V$, 
and the condition at the potential minimum $\eta(x_m)=0$. 
Integration leads to the electric-field distribution

\begin{eqnarray} \label{e}
E = - \frac{d\eta}{dx}=
\left\{ \matrix{
\phantom{+}\sqrt{2 n_m h_V^-(\eta)}, & 0<x<x_m, \cr
-\sqrt{2 n_m h_V^+(\eta)}, & x_m<x< \lambda, } \right. \\
h_V^{\mp}(\eta) = e^{\eta}-1
\pm \beta \left(e^{\eta}{\rm erf}\sqrt{\eta}
- {2\over \sqrt{\pi}}\sqrt{\eta}\right), \label{h}
\end{eqnarray}
where $\beta$ and $n_m$ are functions of $V$ as specified above.
Integrating Eq.\ (\ref{e}), one obtains the distribution of the potential
in an implicit form where, for the given $V$, $\lambda$, the only unknown
parameter is the potential minimum $V_m$. 
The latter is found from the matching at $x=x_m$

\begin{equation} \label{Vm}
\lambda\sqrt{2n_m} =\int_0^{V_m} \frac{d\eta}{\sqrt{h_V^-(\eta)}}
+ \int_0^{V_m+V} \frac{d\eta}{\sqrt{h_V^+(\eta)}}.
\end{equation}
This brief description of the steady-state is then completed by the expression
for the stationary current $I=qA\int v_x F dv_x$, for which we find
\cite{remark2}

\begin{equation} \label{j}
I = I_c e^{-V_m} [1 - e^{-V}] = 2I_c n_m\beta,
\end{equation}
where $I_c=\frac{1}{\sqrt{\pi}}q N_0 v_0 A$ is the emission current from
each contact. 
The above relations solve completely the steady-state problem for a ballistic
conductor under a space-charge-limited transport regime,
for which Eqs.~(\ref{Vm}) and (\ref{j}) determine the current-voltage
characteristics.

To solve the fluctuation problem (\ref{vlasf})--(\ref{dpois}), we first
find the fluctuation of the distribution function $\delta F$ in 
a given electrostatic potential $\varphi(x)+\delta\varphi(x)$
by solving the perturbed kinetic equation (\ref{vlasf}).
The fluctuation $\delta F$ consists of the contributions corresponding to the 
transmitted and reflected groups of carriers, the expressions for them are 
quite cumbersome and will be presented elsewhere \cite{unpub}.
Then, for the current fluctuation we obtain \cite{remark2}

\begin{eqnarray} \label{dj}
\delta I = \int_{V_m}^{\infty} \delta I_L(\varepsilon)\,d\varepsilon
- \int_{V_m+V}^{\infty} \delta I_R(\varepsilon) \,d\varepsilon
- I \delta V_m,
\end{eqnarray}
where $\delta V_m\equiv -\delta\psi(x_m)$ is the potential minimum fluctuation 
and $\delta I_k(\varepsilon)$ is the fluctuation of the contact injection 
current per energy interval 
[$\propto \delta F_k(\varepsilon)$], with $\varepsilon$
being the kinetic energy normalized by $k_B T$.
The correlator for $\delta I_k$ is obtained from that for $\delta F_k$ 
given above, 
$\langle \delta I_k(\varepsilon)\delta I_{k'}(\varepsilon') \rangle 
= 2qI_c (\Delta f) e^{-\varepsilon}\delta_{kk'}
\delta (\varepsilon-\varepsilon')$, 
with $\Delta f$ the frequency bandwidth.
Since the injected electrons of different energies are uncorrelated, 
the first two terms in rhs of Eq.\ (\ref{dj}) give the full shot noise. 
It is the last term $-I\delta V_m$, caused by the self-consistent potential 
fluctuation (long-range Coulomb correlations), 
that compensates the current fluctuation and may result in the noise reduction.

The potential barrier fluctuation $\delta V_m$, which is of prime interest, 
we find from the linearized Poisson equation (\ref{dpois}). 
Let us introduce the perturbation of the potential 
$\delta\eta_x=\delta\psi(x)-\delta\psi(x_m)$, 
referenced to the {\em fluctuating} potential minimum, 
so that at the minimum $\delta\eta_{x_m}$=0.
Thus, Eq.\ (\ref{dpois}) gives the stochastic differential equation 

\begin{equation} \label{nhom}
\frac{d^2\delta\eta_x}{dx^2} = n(x)\, [\delta\eta_x - \delta\eta_0]
\pm \frac{J\,\delta\eta_x}{\sqrt{4\pi\eta(x)}} + \delta n_x^{inj},
\end{equation}
subject to the boundary conditions
$\delta\eta_0$=$\delta\eta_{\lambda}$=$\delta V_m$ \cite{remark3}, where
$J\equiv I/I_c$ is the normalized current and 

\begin{eqnarray} \label{dninjtot}
\delta &&n_x^{inj} = \frac{1}{2\sqrt{\pi}I_c}
\sum_{k=L,R} \int_{\psi_k-\psi_m}^{\infty}
\frac{\delta I_k(\varepsilon)\,
d\epsilon}{\sqrt{\varepsilon+\psi(x)-\psi_k}}
\nonumber\\ &&+ \frac{1}{\sqrt{\pi}I_c} \left\{ \matrix{ 
\int_{\psi_L-\psi(x)}^{\psi_L-\psi_m} \frac{\delta I_L(\varepsilon)\,
d\epsilon}{\sqrt{\varepsilon+\psi(x)-\psi_L}}, & 0<x<x_m, \cr 
\int_{\psi_R-\psi(x)}^{\psi_R-\psi_m}
\frac{\delta I_R(\varepsilon)\,
d\epsilon}{\sqrt{\varepsilon+\psi(x)-\psi_R}}, & x_m<x< \lambda, } \right.
\end{eqnarray}
is the electron-density fluctuation due to the injection from the contacts, 
which is obtained by considering the contributions from both the transmitted 
and reflected groups of carriers.
The terms $\sim\delta\eta_x$ in the rhs of Eq.\ (\ref{nhom}) are related to
the fluctuations of the potential profile induced by injected electrons. 
Note that Eq.\ (\ref{nhom}) is a second-order nonhomogeneous differential
equation with spatially dependent coefficients.
To find its solution in a general form is a complicated problem.
An additional difficulty is due to the term 
$1/\sqrt{4\pi\eta(x)}$ which is singular at $x=x_m$.
Nevertheless, we solve it analytically without any approximation
by making use of the method recently applied 
for a stochastic drift-diffusion equation \cite{apl97},
%one can find the potential fluctuation $\delta\eta_x$ at any section $x$
%of the sample. In particular, 
%It allows us to express the potential barrier fluctuation through those 
%at the contacts
and obtain

\begin{eqnarray}
\delta &&V_m = \frac{1}{\Delta}
\int_0^{\lambda} u(x) \delta n_x^{inj}\,dx, \label{dvmd} \\
\label{u}
u&&(x)= \frac{1}{n(x)} + E(x) \times \nonumber\\ && \left\{ \matrix{
\int_0^x \frac{J\nu(y)+n(y)}{n^2(y)}\,dy -\frac{1}{n_L E_L}, 
& 0 < x < x_m, \cr 
\int_x^{\lambda}\frac{J\nu(y)-n(y)}{n^2(y)}\,dy -\frac{1}{n_R E_R}, 
& x_m < x < \lambda, } \right.
\end{eqnarray}
where $\Delta\equiv (\lambda/2) + E_L^{-1} - E_R^{-1}$,
$\nu(x)\equiv 1/\sqrt{4\pi\eta(x)}$, and 
$n(x)$, $E(x)$ are the steady-state spatial profiles of the electron density 
and electric field, which take the values at the left and right contacts 
$n_L,E_L$ and $n_R,E_R$, respectively. 
The obtained analytical expressions yield the fluctuation of
the barrier height in terms of the spatially distributed ``noise source''
$\delta n_x^{inj}$ which, in turn, is given by the fluctuations 
$\delta I_L$, $\delta I_R$ at the contacts.
The function $u(x)$ shows the relative contributions of the ``noise sources'' 
to the potential barrier fluctuations. 
Substituting the obtained formula for $\delta V_m$ into Eq.\ (\ref{dj}), 
we obtain the current fluctuation as

\begin{eqnarray}
&&\delta I = \int_0^{\infty}
\gamma_L(\varepsilon)\delta I_L(\varepsilon)\,d\varepsilon +
\int_0^{\infty}\gamma_R(\varepsilon)\delta I_R(\varepsilon)\,d\varepsilon,
\label{dJg} \\
&&\gamma_L(\varepsilon) = \left\{ \matrix{
-2 J \int_0^{x_L^*}
K(x,\varepsilon)\,dx, & \varepsilon < V_m, \cr
1 - J \int_0^{\lambda}
K(x,\varepsilon)\,dx, & \varepsilon > V_m, }
\right. \label{gam1} \\
&&\gamma_R(\varepsilon) = \left\{ \matrix{
-2 J \int_{x_R^*}^{\lambda}
K(x,\varepsilon-V)\,dx, & \varepsilon < V_m+V, \cr
-1 - J \int_0^{\lambda}
K(x,\varepsilon-V)\,dx, & \varepsilon > V_m+V, } \right. \label{gam2}
\end{eqnarray}
where
$K(x,\varepsilon)=u(x)/[2\sqrt{\pi}\Delta\sqrt{\varepsilon+\psi(x)}]$,
and $x_L^*$, $x_R^*$ are found from 
$\varepsilon$=$-\psi(x_L^*)$=$V-\psi(x_R^*)$.
The functions $\gamma_k(\varepsilon)$ introduced here for each contact
have a meaning of the {\em current fluctuation transfer functions},
since they represent the ratio of the transmitted current fluctuation to the
injected current fluctuation for a particular injection energy $\varepsilon$.
The terms proportional to the current $J$ are originated from the potential 
minimum fluctuations, whereas the constant contributions ($\pm 1$) represent 
the direct transmission of fluctuations to the opposite 
contact.
Eq.\ (\ref{dJg}) leads to the spectral density of current fluctuations

\begin{equation} \label{SI}
S_I = 2qI_c \int_0^{\infty} \left[ \gamma_L^2(\varepsilon)
+ \gamma_R^2(\varepsilon) \right] e^{-\varepsilon}\,d\varepsilon. 
\end{equation}
This equation with $\gamma_k(\varepsilon)$ given by formulas
(\ref{gam1}) and (\ref{gam2}) is the final result of our derivations.
It allows us to obtain the current-noise spectral density, for the given 
length of the conductor $\lambda$ and applied voltage $V$, from
{\em the steady-state distributions} of the potential $\psi(x)$, electric 
field $E(x)$, and electron density $n(x)$ by direct integration.
Thus, the current-noise level is directly related to the transport 
inhomogeneity in the system.
Note that the obtained formulas are exact for biases ranging from
thermal to shot noise limits under a space-charge-limited transport
conditions. 

A great advantage of the derived formulas is that one may estimate a relative 
contribution to the noise from different energy groups of carriers.
Indeed, Fig.\ 1 shows the functions $\gamma_k(\varepsilon)$ for a fixed 
$\lambda$ and various biases $V$. 
In the low-voltage limit, $\gamma_k$ tend to the step functions with a step 
at the barrier height. 
This means that only electrons able to pass over the barrier contribute 
to the equilibrium (thermal) noise. 
For this case, one can easily obtain the Nyquist noise formula
$S_I^{eq} = 4qI_c e^{-V_m}=4k_BT\,G_0$,
where $G_0=dI/dU|_{U\to 0}$ is the zero-bias conductance.
With increasing the bias $V$, all electrons contribute to the noise:
those transmitted over the barrier, and those reflected back to the contacts. 
The electrons for which $\gamma_k(\varepsilon)<0$ reduce the current 
fluctuations. 
The most efficient in such a compensation carriers are those with the energies 
in the vicinity of the potential barrier energy, where $\gamma_k\to -\infty$
\cite{remark4}. They provide an over-compensation of the injected from 
the contacts fluctuation. 
There also exist the specific energy $\varepsilon^*$, 
for which the compensation fluctuation is exactly equal to the injected 
fluctuation, giving no noise at all $\gamma_L(\varepsilon^*)$=0.

The obtained current-noise spectral density $S_I$, which accounts for 
the long-range Coulomb correlations, may be compared with the uncorrelated 
value through the so-called noise reduction factor 
$\Gamma=S_I/[2qI\coth(V/2)]$. 
By this definition, both the thermal noise and shot noise limits are included 
\cite{gonzalez97}. 
The results (both the noise and the steady-state spatial profiles) 
are in excellent agreement with the Monte Carlo simulations \cite{gonzalez97}. 
Fig.\ 2 shows $\Gamma$ vs applied voltage $V$. 
At low values of $\lambda$, $\Gamma\approx 1$.
As $\lambda$ increases, the noise level becomes substantially reduced
at $k_B T\lesssim qU < qU_{cr}$, where $U_{cr}$ is a critical voltage
for which the potential minimum vanishes. 
At $U\ge U_{cr}$ the full shot noise level is abruptly recovered. 
This sharp increase in the noise intensity when observed in an experiment
would indicate on the disappearance of the potential barrier controlling 
the current.

The obtained exact solutions allows us to investigate in great detail 
the correlations between different groups of carriers. 
While the injected carriers are uncorrelated, those in the volume of the 
conductor become strongly correlated. 
Those correlations may be observed experimentally by making use of 
a combination of two already realized techniques: a hot-electron 
spectrometer \cite{hayes85,heiblum85} and shot-noise measurements 
\cite{reznikov95,kumar96,schoel97}. 
The electron spectrometer, placed 
behind the receiving semitransparent contact, acts as an analyzer 
of electron distribution over the energy \cite{hayes85,heiblum85}. 
In this way a spectroscopic information, that is the average partial currents
$I(\tilde{\varepsilon})$ and their fluctuations 
$\delta I(\tilde{\varepsilon})$ 
may be measured for different energies $\tilde{\varepsilon}$ of electrons 
collected at the contact. 
The partial current of transmitted electrons at the receiving (right) contact 
is given by $I(\tilde{\varepsilon})=
I_c e^{-\tilde{\varepsilon}-V_m}\theta(\tilde{\varepsilon})$, where
$\theta$ is the Heaviside step function, and the threshold energy 
$\tilde{\varepsilon}$=0 corresponds to those arriving electrons 
which have a zero longitudinal kinetic energy at the potential minimum. 
The fluctuation of the partial current is obtained as

\begin{equation} \label{diout}
\delta I(\tilde{\varepsilon}) =
\delta I_L(\tilde{\varepsilon}+V_m)\, \theta(\tilde{\varepsilon})
- I_c\,e^{-V_m}\,\delta V_m\,\delta(\tilde{\varepsilon}).
\end{equation}
Thus, the correlation function 
$\langle\delta I(\tilde{\varepsilon})\,\delta I(\tilde{\varepsilon}')\rangle
|_{x=\lambda}$ for outcoming electrons may be expressed through that for
injected uncorrelated electrons.
A simple analysis shows that for 
$\tilde{\varepsilon}, \tilde{\varepsilon}'>0$ 
the outcoming carriers remain uncorrelated 
since 
$\langle\delta I_L(\tilde{\varepsilon})\,
\delta I_L(\tilde{\varepsilon}')\rangle 
\propto\delta(\tilde{\varepsilon}-\tilde{\varepsilon}')$ due to 
the imposed injection conditions leading to the full shot noise. 
In such a case, an interesting question arises:
what is the reason for the noise reduction obtained for the total 
(integrated over the energies) current fluctuations? 
The answer is found looking at the electrons with energies close to the
threshold energy $\tilde{\varepsilon}$=0 (''tangent'' electrons). 
All other electrons are anticorrelated with that group. 
This means that if there is a positive fluctuation of over-barrier electrons, 
there should be a negative one for the ''tangent'' electrons and vice versa. 
This anticorrelation explains the overall noise reduction. 
The "tangent" electrons can be thought as over-correlated. 
The dispersion $\langle\delta I^2(\tilde{\varepsilon})\rangle$ 
has a sharp peak at $\tilde{\varepsilon}$=0 and then decreases with energy 
at $\tilde{\varepsilon} > 0$. 
This peak is divergent ($\delta$-shaped) in our collisionless theory. 
A small probability of scattering will lead to its broadening 
and finite magnitude. 
Therefore, by measuring the dispersion of the partial current fluctuations 
and/or their cross-correlations, one may observe a 
sharp peak and anticorrelation of electrons, 
thus making the Coulomb correlations effect visible. 

In conclusion, by solving analytically the kinetic equation coupled 
self-consistently with a Poisson equation, 
we have derived a general formula for the current noise in a ballistic 
nondegenerate conductor which accounts for the Coulomb correlations. 
We propose an evident experiment to discover these correlations 
by monitoring different electron energy groups. 
Our work then offers new perspectives on what concerns the study 
of Coulomb interaction and noise in small-size ballistic devices, 
like ballistic transistors, point contacts, etc. 

This work has been supported by 
the Direcci\'on General de Ense\~nanza Superior, Spain 
and the NATO linkage grant HTECH.LG 974610.

\narrowtext

\begin{figure}
{\epsfxsize=7.0cm\epsfysize=6.0cm\centerline{\epsfbox{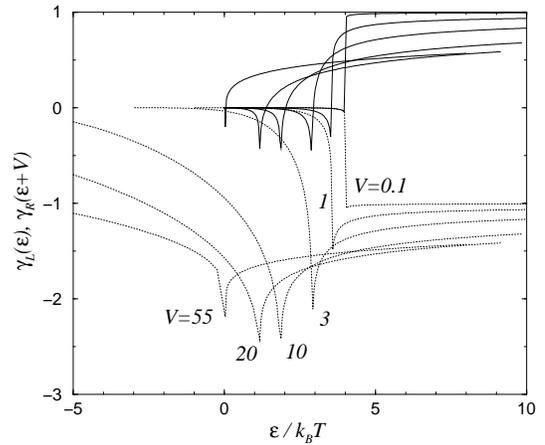}}}
\protect\vspace{0.5cm}
\caption{Current fluctuation transfer functions for each injecting contact 
$\gamma_L(\varepsilon)$ (solid), $\gamma_R(\varepsilon+V)$ (dots)
vs injecting energy $\varepsilon$ for $\lambda$=30 and various biases $V$. 
The argument of $\gamma_R$ is shifted by $V$, so that both $\gamma_k$ 
are singular ($\gamma_k\to -\infty$) at the same energy corresponding to 
the barrier height $\varepsilon=V_m$.}
\label{f1}
\end{figure}

\begin{figure}
{\epsfxsize=7.0cm\epsfysize=6.0cm\centerline{\epsfbox{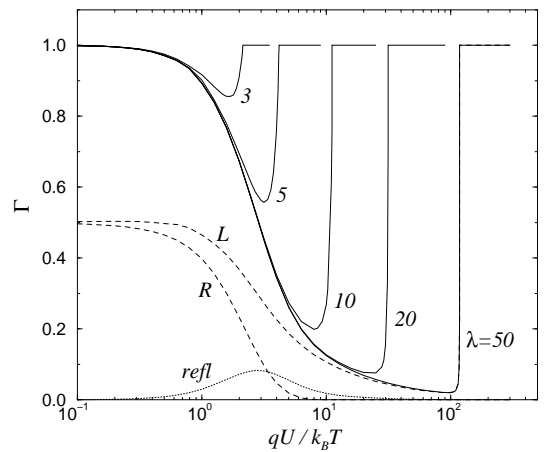}}}
\protect\vspace{0.5cm}
\caption{Current-noise reduction factor $\Gamma$ vs bias $U$
for different lengths of the sample $\lambda$=$d/L_D$. 
For the case of $\lambda$=50, different contributions to $\Gamma$ are shown: 
from over-barrier electrons transmitted from the left (L) and right (R) 
contacts, and those reflected by the potential barrier (refl).}
\label{f2}
\end{figure}

\end{multicols}

\begin{references}
\bibitem{remark1}
Nondegeneracy of an electron gas in our theory doesn't mean that we address
only the case of moderately doped semiconductors.
The theory is also applicable to quantum heterostructures with over-barrier
transport \cite{mitin99}, where current is determined by a tail
in the distribution function 
(ballistic-injection, real-space-transfer devices, etc.).
\bibitem{remark2}
Both the current and its fluctuation are conserved along the sample 
due to the conservation of an electron energy under the ballistic motion.
\bibitem{remark3}
Since the potential $\delta\eta_x$ is referenced to the fluctuating minimum, 
its values on the contacts are not zero, while in a stationary frame 
$\delta\varphi_L$=$\delta\varphi_R$=0 
due to a fixed-applied-voltage conditions.
\bibitem{remark4}
The singularity in $\gamma_k(\varepsilon)$ at $\varepsilon\to V_m$ is
of the logarithmic type, so it gives the well defined net noise after the
integration over $\varepsilon$.
\end{references}
\end{document}